\documentclass[aps,prb,twocolumn,amsfonts,amsmath,amssymb,superscriptaddress]{revtex4}
\usepackage{graphicx}
\usepackage{dcolumn}
\usepackage{bm}

\usepackage{colordvi}
\usepackage{color}

\begin{document}

\title{Spin polarization, dephasing and  photoinduced spin diffusion in (110)-grown  two-dimensional electron systems}

\author{R.\ V\"olkl}
\affiliation{Institut f\"ur Experimentelle und Angewandte Physik,
Universit\"at Regensburg, D-93040 Regensburg, Germany}
\author{M. Schwemmer}
\affiliation{Institut f\"ur Experimentelle und Angewandte Physik,
Universit\"at Regensburg, D-93040 Regensburg, Germany}
\author{M. Griesbeck}
\affiliation{Institut f\"ur Experimentelle und Angewandte Physik,
Universit\"at Regensburg, D-93040 Regensburg, Germany}
\author{S. A. \ Tarasenko}
\affiliation{A. F. Ioffe Physical-Technical Institute, Russian Academy of Sciences, 194021 St. Petersburg, Russia}
\author{D.\ Schuh}
\affiliation{Institut f\"ur Experimentelle und Angewandte Physik,
Universit\"at Regensburg, D-93040 Regensburg, Germany}
\author{W.\ Wegscheider}
\affiliation{Solid State Physics Laboratory, ETH Zurich, 8093 Zurich, Switzerland}
\author{C.\ Sch\"uller}
\affiliation{Institut f\"ur Experimentelle und Angewandte Physik,
Universit\"at Regensburg, D-93040 Regensburg, Germany}
\author{T.\ Korn}
\email{tobias.korn@physik.uni-regensburg.de}
\affiliation{Institut
f\"ur Experimentelle und Angewandte Physik, Universit\"at
Regensburg, D-93040 Regensburg, Germany}
\date{\today}

\begin{abstract}
We study the  optically induced spin polarization, spin dephasing and diffusion  in several high-mobility two-dimensional electron systems, which are embedded in GaAs quantum wells grown on (110)-oriented substrates.  The experimental techniques comprise a two-beam magneto-optical spectroscopy system and polarization-resolved photoluminescence. Under weak excitation conditions at liquid-helium temperatures, we observe spin lifetimes above 100~ns in one of our samples, which are reduced with increasing excitation density due to additional, hole-mediated, spin dephasing. The spin dynamic is strongly influenced by the carrier density and the ionization of remote donors, which can be controlled by temperature and above-barrier illumination. The absolute value of the average electron spin polarization in  the samples is directly observable in  the circular polarization of photoluminescence  collected under circularly polarized excitation and reaches values of about 5~percent. Spin diffusion is studied by varying the distance between pump and probe beams in  micro-spectroscopy experiments. We observe diffusion lengths above 100~$\mu$m and, at high excitation intensity, a nonmonotonic dependence of the  spin polarization on the pump-probe distance.
\end{abstract}

\maketitle
\section{Introduction}
The study of spin dynamics, spin dephasing and spin diffusion is an essential part of semiconductor spintronics~\cite{awschalom02,fabian07,WuReview} research, which ultimately aims at  the implementation of the spin degree of freedom in micro- and nano-electronic devices~\cite{DattaDas,Schliemann03,kunihashi:113502, Koo09182009}. Recently, applications combining optics and semiconductor spintronics have been suggested and demonstrated experimentally: the injection of spin-polarized carriers into semiconductor lasers reduces the laser threshold~\cite{Oestreich_SpinLaser} and can allow for rapid amplitude modulation of the laser emission~\cite{Zutic_SpinLaser,Hofmann_SpinLaser}. It was also shown  that the combination of a spin-polarized  electron gas with a microcavity may yield very large Faraday rotation angles, with possible applications in fast light modulation devices~\cite{Bloch_GiantFaraday}.

The optical orientation and spin dynamics of free carriers in direct-gap semiconductors such as GaAs have been studied intensively.  In n-doped  bulk GaAs  and two-dimensional electron systems (2DES) embedded in GaAs/AlGaAs quantum wells (QWs), electron  spin dephasing at low temperatures is mostly governed by the Dyakonov-Perel (DP) mechanism~\cite{DP}, in which dephasing occurs due to spin precession in the spin-orbit  effective magnetic field in conjunction with momentum scattering. The DP mechanism can be suppressed by changing the symmetry of the spin-orbit field. In two-dimensional structures grown along the [110] crystallographic direction, the Dresselhaus spin-orbit field points along the growth direction. Therefore, the DP mechanism for the electron spin oriented along the growth direction is suppressed in the absence of an additional Rashba spin-orbit field~\cite{DP110}, and spin dephasing is modified if a combination of Rashba and Dresselhaus fields is present~\cite{tarasenko:165317}.  In  symmetric modulation-doped  (110)-grown structures, the spin dephasing times  seem to be limited by the presence of random Rashba fields arising from the inhomogeneous distribution of the remote dopants~\cite{Sherman_RandomRashba,Tarasenko13}.

The spin dephasing in (110)-grown QWs and 2DES was studied experimentally by several groups~\cite{Ohno99_1,ohno01,Oestreich04_1, oestreich08, Voelkl11,Griesbeck12}. Due to the suppression of the DP mechanism, other spin dephasing mechanisms can be studied, yet the measurement of the  lifetimes by optical techniques remains challenging due to spin dephasing induced via optically generated holes.
A number of optical studies on spin diffusion using two-beam Hanle experiments were performed on n-bulk GaAs samples~\cite{awschalom99,crooker05_1,crooker05_2,crooker07,Ossau09}, which show spin diffusion lengths well above 10~$\mu m$. The mobility in these systems, however, is very low due to the direct doping, and they do not allow for easy manipulation of the carrier density and the Rashba spin-orbit interaction by external gate voltages. Optical measurements of spin diffusion in  QWs and 2DES are in many cases more challenging, as the spin diffusion length is shorter due to fast dephasing, so that techniques such as transient spin gratings~\cite{SpinGrating96} or shadow gratings~\cite{Chen:12} need to be applied. Large spin diffusion lengths can be observed in (110)-grown QWs in which carrier transport is facilitated by surface acoustic waves~\cite{Santos_transport110_PRL07, Santos_transport110_PRB08, Santos10}and in (110)-oriented 2DES~\cite{Voelkl11,Marie11}.

Here, we present measurements of spin lifetimes, spin polarization, and spin diffusion  lengths in high-mobility (110)-grown 2DES. We observe long spin lifetimes above 100~ns in the limit of weak optical excitation, and demonstrate that the spin dephasing drastically  depends on the carrier density and the ionization of the modulation doping, which can be controlled by temperature and weak above-barrier illumination. The time-averaged spin polarization degree of  electrons  can be extracted from photoluminescence (PL) measurements and reaches values of several percent.  In experiments with high spatial resolution, we observe spin diffusion lengths above 100~$\mu$m and, for elevated excitation densities, nonmonotonic spin diffusion profiles.

\section{Sample structure and experimental methods}
Samples from three different wafers are studied in our experiment. All three wafers have a similar, complex growth structure, in which a GaAs QW is embedded in AlGaAs barriers, with a total of 4 doping layers placed symmetrically below and above the QW to ensure a near-symmetrical band profile~\cite{Umansky20091658}. In addition to samples from wafer A, which were already investigated previously~\cite{Voelkl11,Griesbeck12}, both, a sample with higher density and similar QW width (sample B), and a sample with a more narrow QW (sample C), were studied.
\begin{table}
\begin{tabular}
{|l|c|c|c|c|c|c|c|c|}
\hline
\# &  width     & density n   & $E_F$ & $\mu$ & $D_n$ &  $D_z^{({\rm ee})}$  & $|g|$  \\
 & nm       &  $\frac{10^{11}}{\textrm{cm}^2}$ & meV & $\frac{10^{6}\textrm{cm}^2}{\textrm{Vs}}$ & $\frac{10^{3}\textrm{cm}^2}{\textrm{s}}$ & $\frac{\textrm{cm}^2}{\textrm{s}}$ & \\
\hline
A           & 30     & 2.7   & 9.6 & 2.14 & 20.5 & 50 (35~K) & 0.38     \\     
\hline
B          & 30      & 3.3   & 11.8 & 3.95  & 46.6 & & 0.37     \\ 
\hline
C          & 20      & 1.2    & 4.3 & 0.74   & 3.2 & 11 (55~K)& 0.32   \\ 
\hline
\end{tabular}
\caption{Characteristic properties of the samples studied. Densities and mobilities
have been determined from magnetotransport measurements at 1.5~K. The spin and charge diffusion coefficients, $D_z^{({\rm ee})}$  and $D_n$,  were calculated using equations~(\ref{Diff_spin}) and~(\ref{Diffcharge_EF}), respectively.  For calculation of $D_z^{({\rm ee})}$, the electron-electron scattering time \ref{ee2D}) and carrier densities for the temperatures corresponding to the experimental conditions in the spin diffusion measurements were used. In sample C, the carrier density $n$ increases with temperature up to $n = 2\cdot 10^{11}\textrm{cm}^{-2}$ above 20~K due to full ionization of the remote dopants. The g factors were measured by time-resolved Faraday and Kerr rotation at liquid-helium temperature, changes of the g factors can be neglected in the temperature range studied here.} \label{Data}
\end{table}

For the measurements, sample pieces measuring 4~mm by 5~mm are cleaved from a wafer. They are mounted in vacuum on the cold finger of a He-flow cryostat.  A tunable Ti-Sapphire continuous-wave (cw) laser is used as a source for the circularly polarized pump beam. It is tuned to $\lambda_x=760$~nm for all measurement series presented here. Therefore, it  nonresonantly excites spin-polarized electron-hole pairs in the QW  with an excess electron energy of about 100~meV, depending on the QW width and the sample temperature.  In order to avoid the buildup of a dynamic nuclear polarization during the experiments, the helicity of the circularly polarized pump beam is modulated at a frequency of 23~Hz using a liquid crystal retarder.  The $z \parallel [110]$ component of the spin polarization is detected by near-resonant probing with a tunable cw diode laser, which is linearly polarized, via the magneto-optic Kerr effect (MOKE). Both beams are superimposed on each other at a beamsplitter and either coupled into   microscope objectives with 20x or 10x magnification  or focussed onto the sample using an achromatic lens with 50~mm focal length. The reflected probe beam is spectrally filtered using a bandpass to suppress the collinear pump beam and then coupled into an optical bridge, which detects the small rotation of the probe beam polarization axis due to the polar MOKE. A lock-in modulation scheme is used to increase the sensitivity of the detection. The spot sizes of pump and probe beams are about 4~$\mu$m, if the 20x microscope objective is used, and about 8~$\mu$m for the 10x objective.  The lens leads to larger spot sizes of 40~$\mu$m for both beams.  The spot size is determined by scanning the beam over a lithographically defined structure. For spatially resolved measurements, the pump beam is scanned with respect to the probe beam by a piezo-controlled mirror. In order to extract spin diffusion profiles, measurements are performed in which the amplitude of the MOKE signal is measured for zero applied field and  the in-plane field of 20~mT, instead of measuring a full Hanle curve for each distance between pump and probe spots. The signal amplitude is determined from the difference of the two measurements.
The cryostat is mounted between a pair of Helmholtz coils, and magnetic fields up to 30~mT can be applied in the sample plane.

For some measurements, an additional, weak above-barrier illumination with a 532~nm cw laser is used. This laser is focused to a large spot diameter which covers the whole sample. To determine the effects of the above-barrier illumination and to measure the  spin polarization degree in the samples, photoluminescence (PL) measurements are performed using slight modifications of the experimental setup described above. Here, only the  pump beam tuned to $\lambda_x=760$~nm is used to create electron-pairs in the sample, it is focused onto the sample using the achromatic lens, so that a large focal spot of 40~$\mu$m is illuminated. Additionally, above-barrier illumination can be used in a similar way as described above. The resulting PL is collected in backscattering geometry using the same lens, coupled into a spectrometer and recorded using  a liquid-nitrogen cooled charge-coupled device (CCD) sensor. To determine the spin polarization via PL measurements, the pump beam is circularly polarized. The circular polarization  of the PL is analyzed by using an achromatic wave plate and a polarizer. The circular polarization degree $P(E)$ is determined from two subsequent measurements, in which the PL in co- and contracircular helicity to the excitation is collected. It is calculated as a function of the PL energy by dividing the difference of the PL intensities $I_{PL}(E)$ for co- and contracircular helicity by their sum.

\section{Theoretical approach}
\subsection{Optical orientation and spin dephasing}
The nonresonant excitation of a QW with circularly polarized light creates spin-polarized electron-hole pairs. While the holes typically lose their spin orientation during momentum relaxation, energy relaxation in the conduction band is mostly spin-conserving.  However, due to valence-band mixing, the spin polarization degree for the optically oriented electrons is not 100~percent under nonresonant excitation conditions, but significantly lower. For excess energies as used in our experiments, an initial spin polarization degree for optically oriented electrons of about 30~percent was observed in a 20~nm wide QW~\cite{oestreich05_1}.
At zero magnetic field  and homogeneous photoexcitation,  the $z$ component of the  electron spin density is given by
\begin{equation}\label{S_Z}
S_z(0) = G_z \tau_z \:,
\end{equation}
 where  $G_z$ is the spin generation rate, proportional to the excitation  density $I$, and $\tau_z$ is the lifetime of electron spin oriented along the QW normal. The decay rate $1/\tau_z$  is determined by three  contributions,
\begin{equation}\label{tau_sz}
1/\tau_{z} = 1/\tau_{z}^{lim} + \gamma_z^{BAP} N_h + \gamma^{r} N_h \:,
\end{equation}
with $1/\tau_{z}^{lim}$  being  the spin dephasing rate in the limit of zero excitation. In the absence of a global Rashba field arising from incomplete and asymmetric ionization of remote dopants, the time $\tau_{z}^{lim}$ is likely to be determined by  the DP mechanism caused by small random Rashba  fields present in any modulation-doped structure~\cite{Sherman_RandomRashba,Wu_randomRashba}. The terms $\gamma_z^{BAP} N_h$ and $\gamma^{r} N_h$ describe the spin decay due to the Bir-Aronov-Pikus (BAP) mechanism~\cite{BAP,APT83}, which is proportional to the steady-state hole density $N_h$, and the recombination of optically oriented electrons with holes, respectively. The role of photocarrier recombination is as follows: in the limit of high excitation density, nearly all of the optically oriented electrons  would recombine with holes, similar to an undoped QW, and the photocarrier lifetime limits the spin lifetime.
It follows  from Eqs.~(\ref{S_Z}) and~(\ref{tau_sz}) that $S_z(0)$ is given by
\begin{equation}\label{S_zerofield2}
S_z(0) = \frac{G_z}{1/\tau_{z}^{lim} + (\gamma_z^{BAP} + \gamma^{r}) N_h}  \:.
\end{equation}
In the limit of very low excitation density, where  the time $\tau_{z}^{lim}$ determines the spin lifetime,  $S_z (0)$ depends linearly on the excitation density $I$, as $G_z \propto I$. For increased excitation density, $S_z$ saturates, as $N_h \propto I$. Due to the low spin dephasing rate $1/\tau_{z}^{lim}$ in absence of a global Rashba field, this saturation can occur at rather low values of the excitation density.  Below we demonstrate that the experimental amplitude data is well-described by the fit function
\begin{equation}\label{S_zerofield3}
S_z(0) \propto \frac{I}{1 + I/I_0}  \:,
\end{equation}
which has the excitation density dependence of  Eq.~(\ref{S_zerofield2}).  We note that, from the excitation-density dependence, the relative magnitude of $\gamma_z^{BAP}$ and $\gamma^{r}$ cannot be determined.  In deriving Eq.~(\ref{S_zerofield3}), we neglect the influence of photoexcitation and excitation-induced heating on the DP mechanism. It will be shown below that this effect  may be important and is pronounced in one of our samples.

The application of an in-plane magnetic field leads to a depolarization of the optically oriented spin polarization due to spin precession.
The dependence of the time-averaged spin  density $S_z$ on the magnetic field is given by the Lorentzian
\begin{equation}\label{Hanle_B}
S_z(B) = \frac{S_z(0)}{1+ (\omega_L \tau_s)^2}.
\end{equation}
Here, $\omega_L=g_e \mu_B B/\hbar$ is the Larmor frequency, which is determined by the electron $g$ factor and the applied magnetic field $B$, and $\mu_B$  is the Bohr magneton. The  in-plane  $g$ factors of our samples have been determined by time-resolved Faraday and Kerr rotation measurements (not shown), the values are given in table~\ref{Data}.
The spin lifetime obtained in Hanle measurements is given by
\begin{equation}\label{tau_s_Hanle}
 \tau_s = \sqrt{\tau_{z} \tau_{\|}} \:,
\end{equation}
where $\tau_{\|}$ is the in-plane spin dephasing time. The rate $1/\tau_{\|}$ can be also presented in the form of Eq.~(\ref{tau_sz}).
However, $1/\tau_{\|}^{lim} \gg 1/\tau_{z}^{lim}$  in  symmetrically (110)-grown structures, as $1/\tau_{\|}^{lim}$ is determined by the conventional DP mechanism in the Dresselhaus field perpendicular to the  QW plane, while $\gamma_{\|}^{BAP}$ and $\gamma_{z}^{BAP}$ are comparable.   The ratio $\tau_z^{lim} / \tau_{\parallel}^{lim} \sim 7$ has been reported previously~\cite{oestreich08}, and even larger anisotropies up to 50 were recently observed in resonant spin amplification (RSA) measurements~\cite{Griesbeck12}.  In the excitation density range we study, we may therefore neglect the effects of BAP mechanism and recombination on the in-plane spin dephasing rate and use the expression for the spin lifetime measured in Hanle experiments
\begin{equation}\label{tau_s_Hanle4}
1/\tau_{s} = \sqrt{1/\tau_{\|}^{lim}} \sqrt{[1/\tau_{z}^{lim} + (\gamma_z^{BAP} + \gamma^{r}) N_h]} \:.
\end{equation}
For $N_h \propto I$, the spin dephasing rate extracted from the Hanle curves has the following intensity dependence:
\begin{equation}\label{tau_s_Hanle5}
1/\tau_{s}(I) = 1/\tau_{s}(0) \sqrt{ 1 + I/I_0} \:.
\end{equation}

At low excitation intensity, the width of the Hanle curve is determined by $\sqrt{\tau_z^{lim}\tau_{\parallel}^{lim}}$.
For the DP mechanism of spin dephasing in the collision-dominated regime, the time $\tau_{\parallel}^{lim}$ is given by
\begin{equation}\label{tau_plane_lim}
1/\tau_{\parallel}^{lim} = \frac{\gamma^2 \langle k_z^2 \rangle^2 \, m^* \tau_p^* \, \tilde{\varepsilon}}{\hbar^4} \:,
\end{equation}
where $\gamma$ is the bulk Dresselhaus constant, $m^*$ is the effective mass, $\tau_p^*$ is the scattering time, $\tilde{\varepsilon} = E_F/[1-\exp(-E_F/k_BT)]$ is a characteristic energy equal to $E_F=\pi n \hbar^2/m^*$ and $k_B T$ for the degenerate distribution with the Fermi energy $E_F$ and Boltzmann distribution with the temperature $T$, respectively.

The time $\tau_p^*$  describes the decay of the first angular harmonic of the spin distribution function in $\bm{k}$-space~\cite{DP110,Harley07}. It determines the spin diffusion coefficient~\cite{Damico00} and the relaxation time of pure spin current~\cite{Ivchenko08}. The time $\tau_p^*$ is limited by, both, electron scattering from static defects and phonons as well as electron-electron collisions. The latter do no directly influence the electron gas mobility but affect the spin dynamics as they lead to an isotropization of the spin distribution in $\bm{k}$-space. This influence is directly observable, e.g., in the decay of coherent precession of spin-polarized electrons~\cite{Shields02,Griesbeck09}. The electron-electron scattering time in a degenerate 2DES with the Fermi energy $E_F$ at the temperature $T \ll E_F/k_B$ can be calculated for the strict two-dimensional limit~\cite{Harley07}
\begin{equation}\label{ee2D}
\frac{1}{\tau_{ee}} \sim 3.4 \frac{E_F}{\hbar} \left( \frac{k_B T}{E_F} \right)^2 .
\end{equation}
In high-mobility 2DES, even at liquid-Helium temperatures, $\tau_{ee}$ is significantly smaller than the momentum relaxation time $\tau_p$, which can be determined from the mobility, and provides the upper limit for the time  $\tau_p^*$.

\subsection{Spin diffusion}
Now we consider spatially inhomogeneous optical excitation and diffusion of the $z$ component of electron spin in the QW plane at zero magnetic field. The  spatial distribution of the spin density induced by the spin pumping $G_z(\bm{r})$ can be found from the diffusion equation
\begin{equation}\label{diffusion}
S_z(\bm{r}) - \tau_z \nabla \cdot [D_z \nabla S_z(\bm{r})] = G_z(\bm{r}) \tau_z  \:,
\end{equation}
where $\bm{r}$ is the in-plane coordinate and $D_z$ is the diffusion coefficient for the $z$ component of the electron spin,
\begin{equation}\label{Diff_spin}
D_z= \frac{\tilde{\varepsilon} \, \tau_p^*}{m^*} \:.
\end{equation}
As discussed above, due to electron-electron collisions the spin diffusion coefficient $D_z$ is significantly smaller than the charge diffusion coefficient $D_n$, which can be obtained from the mobility by using the Einstein relation,

\begin{equation}\label{Diffcharge_EF}
D_n=\mu_n \tilde{\varepsilon} / e \:,
\end{equation}
where $e$ is the elementary charge.

This  reduction of spin diffusion is known as the spin-Coulomb-drag~\cite{Damico00}. The reduction  is  suppressed  at  high electron spin polarization,  where the carrier majority with a certain spin projection determines transport properties. Values for $D_n$ calculated for our samples using the parameters extracted from magnetotransport data are given in table~\ref{Data}. Due to the high carrier mobility in our 2DES, the  charge diffusion coefficients are large.

The solution of the diffusion Eq.~(\ref{diffusion}) for a point-like excitation spot  and spatially-independent $\tau_z$ and $D_z$ has the form
\begin{equation}\label{MacDonald}
S_z(r) = C K_0(r/L_z) \:,
\end{equation}
where $K_0$ is the modified Bessel function of the second kind (MacDonald function), $C$ is a constant depending on the
excitation intensity, and $L_z$ is the spin diffusion length,
\begin{equation}\label{Difflength}
L_z=\sqrt{D_z \tau_z} \:.
\end{equation}
 At $r \gg L_z$, the function $S_z(r)$ has the asymptotic behavior
\begin{equation}\label{diff_asymp}
S_z(r) \propto \frac{\exp(-r/L_z)}{\sqrt{r/L_z}} \:.
\end{equation}

By fitting experimental data  far from the excitation spot, where the influence of holes on electron spin dynamics is negligible, with Eq.~(\ref{diff_asymp}) we can extract the spin diffusion length $L_z$. We note that the MacDonald function yields larger values of $L_z$ than the simple exponential decay function used in our previous work~\cite{Voelkl11}.

The knowledge of the time $\tau_s$ and the length $L_z$ measured independently in similar  experimental conditions allows us to determine other relevant parameters. Indeed, by combining Eqs.~(\ref{tau_s_Hanle}), (\ref{tau_plane_lim}), (\ref{Diff_spin}), and~(\ref{Difflength}) we obtain

\begin{equation}\label{tau_z_est}
\tau_z = \frac{L_z \Omega_D \tau_s}{v} \:,
\end{equation}
\begin{equation}\label{tau_parallel_est}
\tau_{\parallel} = \frac{v \, \tau_s}{L_Z \Omega_D} \:,
\end{equation}
%
%
\begin{equation}\label{D_z_est}
D_z = \frac{L_z v}{\Omega_D \tau_s} \:,
\end{equation}
with $\Omega_D =\gamma \langle k_z^2 \rangle \sqrt{2 \tilde{\varepsilon} m^*}/\hbar^2$ and $v=\sqrt{2\tilde{\varepsilon}/m^*}$.

\section{Results and Discussion}
\subsection{Spin dephasing}
First, we study the spin dephasing in the samples as a function of experimental parameters. To suppress the effects of  spin diffusion out of the pump spot in these measurements, the experiments are performed using the larger spot size of 40~$\mu$m, with full overlap of pump and probe beams.
\begin{figure}
  \includegraphics[width= 0.5\textwidth]{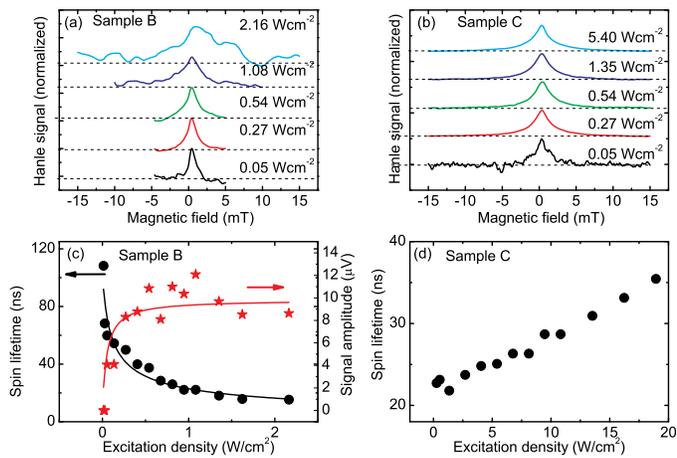}
   \caption{(color online) (a) and (b) Hanle-MOKE traces measured  for different excitation densities on sample B and C at a nominal temperature of 4~K. (c) Spin lifetime $\tau_s$ (filled circles) and MOKE signal amplitude at zero magnetic field (filled stars) for sample B as a function of excitation density. The solid  lines are  fits to  Eq.~\ref{tau_s_Hanle5} (spin lifetime)  and Eq.~\ref{S_zerofield3} (signal amplitude) with $I_0 = 0.052$~W/cm$^2$. (d) Spin lifetime $\tau_s$ for sample C as a function of excitation density. }
   \label{Power_4Panel}
\end{figure}
Figure~\ref{Power_4Panel} (a) and (b) show typical Hanle-MOKE traces measured on sample B and C for different excitation densities of the pump laser. While in sample B, it is clearly visible that the linewidth of the Hanle curve increases significantly with the excitation density, only a small narrowing of the Hanle curve is observed in sample C for an even larger range of excitation densities. The spin lifetimes for both samples, extracted from the linewidth of the Hanle curves, are shown in Figs.~\ref{Power_4Panel} (c) and (d). Here, sample B shows a behavior already observed previously~\cite{Voelkl11} in sample A: the spin lifetime \emph{decreases} with increasing excitation density. For the lowest excitation density, it reaches values of more than 100~ns.  The solid black line in Fig.~\ref{Power_4Panel} (c) corresponds to a fit to the dependence of the spin lifetime using Eq.~(\ref{tau_s_Hanle5}). Additionally, the signal amplitude, which  corresponds to $S_z(0)$ in Eq.~(\ref{Hanle_B}),  saturates at low excitation densities, the solid red line corresponds to a fit to the signal amplitude data using Eq.~(\ref{S_zerofield3}). For both fit curves, the same saturation amplitude $I_0$  $=0.052$ W/cm$^2$ was used.  This radiation intensity corresponds to the photoinduced hole density $\sim 4 \times 10^6$~cm$^{-2}$ for a QW absorbance $\eta \sim 2$~percent, which was determined experimentally for similar QW samples, and a hole lifetime of 1~ns.
The observed behavior may be explained as follows.  The quantum wells in samples A and B are macroscopically symmetric and the BAP mechanism of spin dephasing plays an important role, which leads to a decrease of the spin lifetime $\tau_s$ with increasing the excitation density, see  Eq.~(\ref{tau_s_Hanle5}). The samples differ mainly in the maximum value of  $\tau_s$ that can be reached in the limit of low excitation density, and in the excitation density for which saturation of the signal amplitude occurs.  By contrast, in sample C, the spin lifetime \emph{increases} with the pump excitation density. From this, we infer that the dominant dephasing mechanism for this sample, under the experimental conditions in this measurement series, is the Dyakonov-Perel mechanism.   In this regime, the optically generated holes serve as scattering centers, reducing the momentum relaxation time, and thereby increasing the spin lifetime. Additionally, the increased excitation density leads to local heating, which aids the ionization of remote dopants. We will show below that the dominant dephasing mechanism in sample C changes with temperature.

\begin{figure}
  \includegraphics[width= 0.4\textwidth]{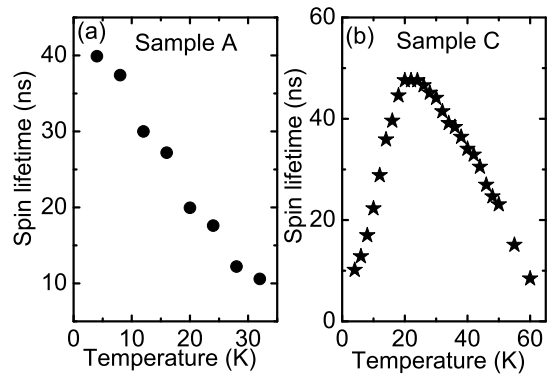}
   \caption{(color online) (a) Spin lifetime as a function of sample temperature for sample A. (b) Spin lifetime as a function of sample temperature for sample C. In both measurement series, an excitation density of 0.27~Wcm$^{-2}$ was used.}
   \label{Tspin_Temp}
\end{figure}

We now look at the temperature dependence of the spin lifetime, which is given in  Fig.~\ref{Tspin_Temp} for samples A and C. While for sample A, the spin lifetime monotonously \emph{decreases} with temperature, sample C shows a pronounced maximum of the spin lifetime at about 20~K. Such a drastic change of the spin lifetime with temperature was recently observed by resonant spin amplification in a similar sample~\cite{Griesbeck12}. The reason for this behavior lies in the complex growth structure of our samples: while the sample design is optimized to yield a highly symmetrical modulation doping, leading to a vanishing Rashba spin-orbit field, we observe that in some wafers, at low temperatures, the remote donors are not fully ionized. This leads to a reduced carrier concentration, but more importantly, it also gives an asymmetric ionization of the dopant layers below and above the quantum well, resulting in a pronounced Rashba field. This leads to spin precession also for the out-of-plane spin  component and makes the DP mechanism dominant. With increasing temperature, the remote dopant layers become fully ionized, increasing the carrier concentration, and also significantly reducing the Rashba field.  Now, the DP mechanism is suppressed for the out-of-plane spin orientation.
The decrease of the spin lifetime with temperature, which is observed in sample A for the whole temperature range investigated, and in sample C for temperatures above 18~K, where ionization of remote dopants is complete, stems from the BAP mechanism.  This mechanism becomes more efficient with increasing temperature because of reduced Pauli blocking and increased electron-hole scattering rates~\cite{Zhou_Wu_BAP}.

Next, we investigate the effects of above-barrier illumination on the spin dynamics. The generation of electron-hole-pairs in the barriers leads to a redistribution of the donor electrons in a two-step process: while optically generated holes in the valence band can easily move from the barrier layers into the QW, there is a potential barrier for electrons in the conduction band which is formed by the modulation doping. The holes may then recombine with electrons in the QW, effectively  transferring electrons from the 2DES back to the dopant layers~\cite{Chaves1986} and reducing the carrier density. This process is also known as optical gating and may even lead to inversion of the carrier type~\cite{syperek07}.
\begin{figure}
  \includegraphics[width= 0.5\textwidth]{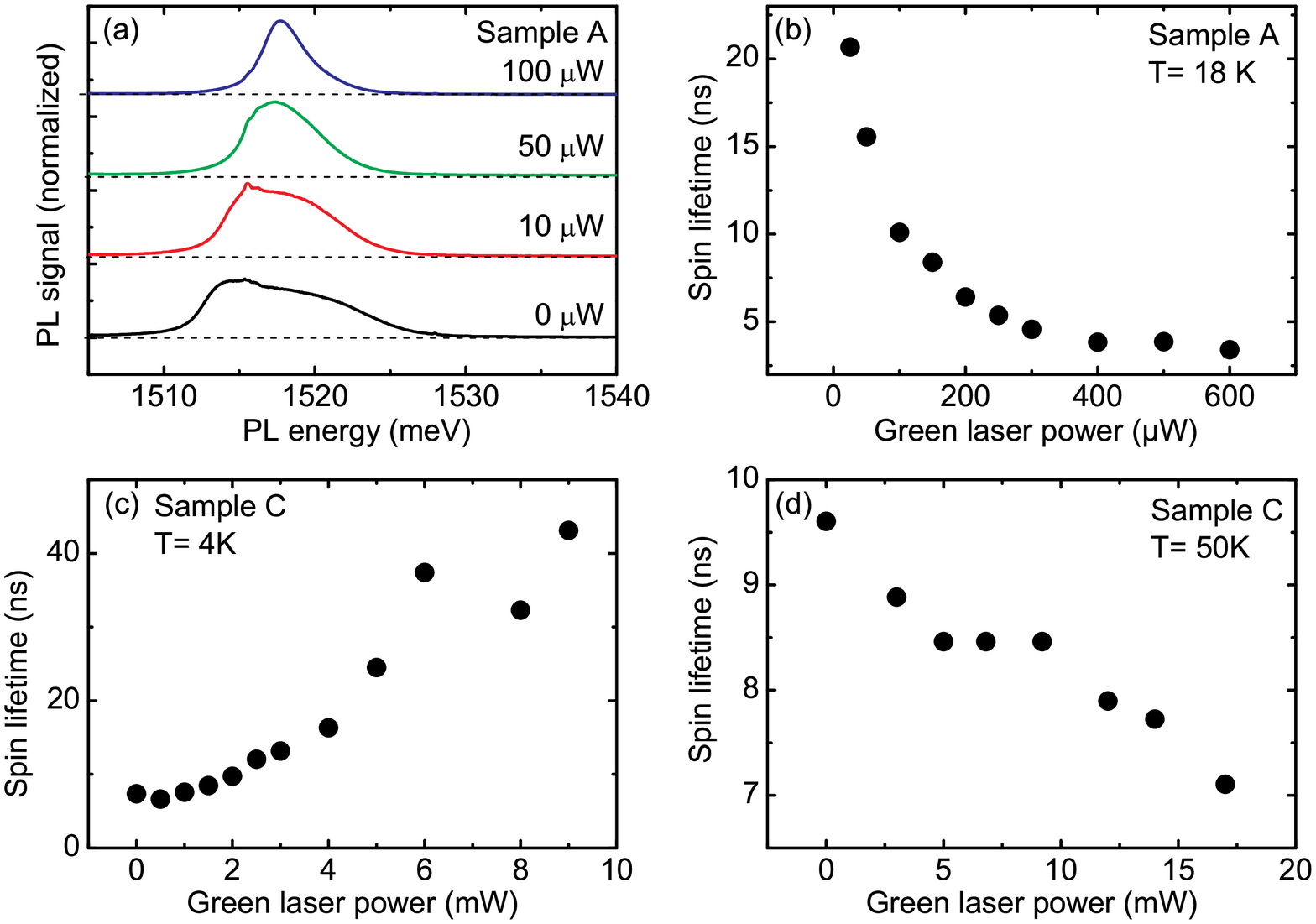}
   \caption{(color online) (a) Photoluminescence spectra of sample A measured for different values of above-barrier illumination power. (b) Spin lifetime as a function of above-barrier illumination power for sample A at a sample temperature of 18~K. An excitation density of 0.27~Wcm$^{-2}$ was used for the pump laser. (c) and (d) Spin lifetime as a function of above-barrier illumination power for sample C at a sample temperature of 4~K (c) and 50~K (d). In both measurement series, an excitation density of 0.54~Wcm$^{-2}$ was used for the pump laser.}
   \label{PL_Green}
\end{figure}
We can directly observe this effect in PL measurements, as shown in Fig.~\ref{PL_Green}(a) for sample A. If no above-barrier illumination is used, the PL from the 2DES shows a typical, shark-fin like shape. The width of the PL peak corresponds to transitions in the QW from the lowest-lying energy states in the conduction band up to the Fermi energy of the 2DES, therefore, the PL width can be used to monitor the carrier density in the 2DES. By increasing the  power of the above-barrier illumination, the width of the PL is reduced significantly. To study the  effect of optical gating  on the spin dynamics, Hanle measurements were performed for a fixed pump intensity, with varying above-barrier illumination power, for samples A and C. The spin lifetimes extracted from the Hanle curves are plotted in Figs.~\ref{PL_Green}(b)-(d)~\footnote{RSA measurements (not shown) demonstrate that the electron g factor changes by about 5~percent in the range of carrier densities that can be accessed by optical gating.}. In sample A, where the BAP mechanism  plays an important role, we see a \emph{decrease} of the spin lifetime with the above-barrier illumination. Here, the reduction of the carrier density in the 2DES reduces Coulomb screening of the photogenerated holes, increasing the electron-hole scattering rate. Therefore, the BAP mechanism becomes more effective, leading to a reduction of the spin lifetime. By contrast, sample C shows a different behavior at low temperatures, evidenced by  Fig.~\ref{PL_Green}(c): here, the spin lifetime \emph{increases} with the above-barrier illumination power. This finding supports the interpretation that, at low temperatures, the spin dephasing in sample C is dominated by the DP mechanism. Therefore, the reduction of the carrier density (and the Fermi energy, accordingly) leads to  a slowdown of the spin dephasing due to  a reduction of the spin-orbit effective magnetic field as well as a decrease of the scattering time $\tau_p^*$. The first effect is caused by the fact that, in QWs, the effective field is proportional to the electron wave vector. The latter results from a reduction of the Coulomb screening and decrease of the electron-electron scattering time $\tau_{ee}$, cf. Eq.~(\ref{ee2D}).

Additionally, the above-barrier illumination of sample C may reduce the asymmetry in the ionization of the remote dopants at low temperatures, leading to a reduction of the Rashba spin-orbit field. This effect was observed recently in RSA measurements on a similar sample~\cite{Griesbeck12}. We note that, at higher temperatures, the dependence of the spin lifetime on the above-barrier illumination changes, as Fig.~\ref{PL_Green}(d) shows. Here, we see a similar behavior as in sample A, indicating that the BAP mechanism becomes dominant in sample C at higher temperatures due to  vanishing of the regular Rashba field in the structure with fully ionized dopants.

\subsection{Spin polarization}
To study the  optically induced spin polarization in our samples, we investigate the circular polarization degree of the PL. In contrast to the Hanle-MOKE measurements, which give only a Kerr rotation angle that is proportional to the time-averaged spin polarization, the circular polarization degree of the PL may be used to determine the absolute spin polarization degree in the sample.
\begin{figure}
  \includegraphics[width= 0.45\textwidth]{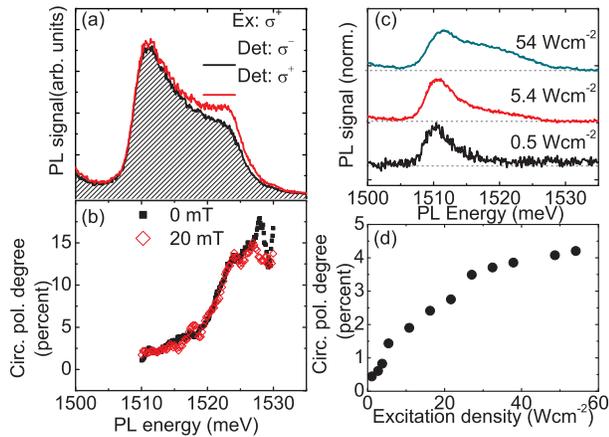}
   \caption{(color online) (a) PL spectra of sample A  measured for co- and contracircular helicity of excitation laser and detection. (b) Circular polarization degree as a function of PL energy extracted from spectra such as shown in (a), with and without an applied magnetic field. It is calculated only for spectral regions with strong PL. An excitation density of 5.4~Wcm$^{-2}$ was used for the measurements shown in (a) and (b). PL spectra of sample A measured as a function of excitation density. (d) Averaged circular polarization degree extracted from PL spectra of sample A as a function of excitation density. }
   \label{PL_PolDeg}
\end{figure}

Figure~\ref{PL_PolDeg}(a) shows two helicity-resolved PL spectra measured on sample A at 4~K using an excitation density of 5.4~Wcm$^{-2}$. It is clearly visible that for co-circular excitation  and detection, there is a larger PL signal observed at the high-energy side of the PL from the 2DES. From the two PL spectra, we can calculate the circular polarization degree $P(E)$ as a function of the PL energy. This is plotted in Fig.~\ref{PL_PolDeg}(b) for the spectral regions in which strong PL is emitted from the sample. We see that the circular polarization degree of the PL is large at the high-energy edge of the PL emitted from the 2DES, reaching almost 20~percent. This observation may be interpreted as follows: in a 2DES at zero temperature in the absence of an external magnetic field, all available states  below the Fermi level are occupied with equal numbers of the spin-up and spin-down electrons, while states above the Fermi level are unoccupied. A spin polarization may only occur if spin-polarized electrons are added to the system above the Fermi energy. Finite temperatures lead to a softening of the Fermi-Dirac function in the region around the Fermi energy, so that a difference in occupation between the spin-up and spin-down electron states is also allowed below the Fermi energy. A similar behavior was observed in highly-doped n-bulk GaAs~\cite{dzhioev02_2}. The electron temperature in the samples during PL measurements has been determined by analyzing the high-energy tail of the PL (see, e.g., ref.~\onlinecite{KornReview} for details) to be about 20~K for a nominal sample temperature of 4.5~K. As an in-plane magnetic field is applied to the sample, depolarization of the PL can occur due to precession of the spin-polarized electrons. The circular polarization degree for the applied field of 20~mT is also depicted in Fig.~\ref{PL_PolDeg}(b). For this magnetic field value,  however, there is no significant change of the circular polarization of the PL as compared to the zero-field measurements, as  for the excitation density  used in the PL measurement the spin lifetime is strongly reduced  via the BAP mechanism.

Given that only PL due to recombination of electrons close to the Fermi energy yields a circular polarization, we  investigate the excitation-density dependence of the PL spectrum in more detail. Figure~\ref{PL_PolDeg}(c) shows a series of PL spectra of sample A measured for different excitation densities.  For low excitation density, the PL  is dominated by the low-energy peak. Under these conditions, the optically generated holes can relax to the valence-band  top during the photocarrier lifetime and, therefore, recombine with electrons close to $k=0$, which do not carry a spin polarization. As the excitation density is increased, optically generated holes also occupy valence-band states with larger $k$ values, so that the PL  spectrum  develops a more pronounced high-energy shoulder due to recombination of spin-polarized electrons close to the Fermi energy. Thus, depending on the excitation density, different subsets of the 2DES are probed in PL experiments.
This is evident in the dependence of the  circular polarization degree (averaged over the PL spectrum) on excitation density, shown in Fig.~\ref{PL_PolDeg}(d). We clearly see that  the average polarization  grows with increasing excitation density, where spin-polarized electrons start to contribute to the PL, and saturates for larger values. The low-excitation-density regime, in which we observe long spin lifetimes and saturation of the MOKE signal amplitude, does not yield a measurable circular polarization degree of the PL.
This finding is an indication that PL-based measurements are of limited use to study the spin dynamics and optical orientation in samples with degenerate electron or hole systems.

\subsection{Spin diffusion}
Now, we study the spin diffusion in our samples as a function of experimental parameters.
\begin{figure}
  \includegraphics[width= 0.5\textwidth]{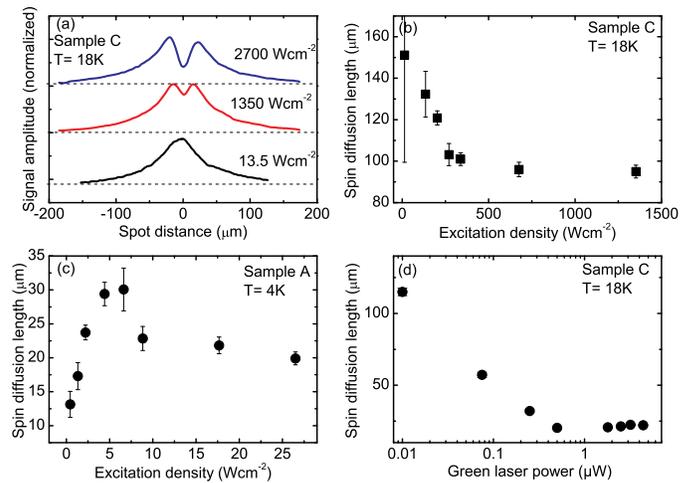}
   \caption{(a) Spin diffusion profiles measured on sample C for different excitation densities. (b) Spin diffusion length in sample C as a function of excitation density. (c) Spin diffusion length in sample A as a function of excitation density. (d) Spin diffusion length in sample C as a function of above-barrier illumination.}
   \label{SpinDiff_4Panel}
\end{figure}
Figure~\ref{SpinDiff_4Panel}(a) shows spin diffusion profiles measured on sample C for different excitation densities at a temperature of 18~K, where the spin lifetime in sample C reaches its maximum [see Fig.~\ref{Tspin_Temp}(b)] due to complete ionization of the remote donors. For the lowest excitation density depicted, the diffusion profile shows a monotonic \emph{decrease} of the time-averaged spin polarization as a function of the distance between the pump and probe spots. A diffusion length of 150~$\mu$m is extracted from this profile by fitting  the experimental data by the MacDonald function~(\ref{MacDonald}). As the excitation density is increased, a pronounced minimum develops at the overlap of pump and probe beams. This behavior was previously reported also for sample A and can be explained as follows: at the overlap of the pump and probe spots, the spin dephasing  speeds up due to the presence of holes generated by the pump beam. The hole density is mostly confined to the pump spot, as the hole diffusion is suppressed by photocarrier recombination and the large  hole effective mass.

Far from the excitation spot, holes are absent and the spin polarization monotonically decays with the distance between the pump and  probe spots. The decay is well fitted by Eq.~(\ref{MacDonald}), which enables us to determine the spin diffusion length $L_z$. The extracted dependence of $L_z$ on the excitation density for samples A and C is shown in Fig.~\ref{SpinDiff_4Panel}.
In sample A, we observed a large increase of the spin diffusion length with increasing excitation density  and then a slight decay, Fig.~\ref{SpinDiff_4Panel}(c). Remarkably, the spin diffusion length in sample C, depicted in Fig.~\ref{SpinDiff_4Panel}(b), does not show such an increase  for the excitation density range studied, but  only a slight decrease.  We attribute the growth and saturation of the spin diffusion length with the excitation density in sample A with the suppression of the spin-Coulomb-drag~\cite{Damico00} at high spin polarization degree.
In sample C, where the resident carrier concentration is low and the spin lifetime is long, spin polarization can be very high in this measurement series, so that an influence of the spin-Coulomb-drag on spin diffusion is not that pronounced and effects of electron gas heating can be more critical.
By contrast, weak above-barrier illumination strongly modifies the spin diffusion profiles in sample C, as demonstrated in  Fig.~\ref{SpinDiff_4Panel}(d). Here, a clear transition from large spin diffusion length values to small values occurs for a critical value of the above-barrier illumination power. This observation indicates that the reduced carrier density leads to a changeover from degenerate to nondegenerate regime in the 2DES, so that it follows Boltzmann statistics. In sample C, PL measurements in combination with above-barrier illumination (not shown) indicate that the carrier density can be reduced to less than 3$\times10^{10}$~cm$^{-2}$, corresponding to a Fermi temperature below 15~K.
This leads to a decrease of the average electron energy, even if it is well above the lattice temperature due to the nonresonant excitation~\cite{Ossau12}, and, hence, to a slowdown of diffusion, see Eq.~(\ref{Diff_spin}).
\begin{figure}
  \includegraphics[width= 0.4\textwidth]{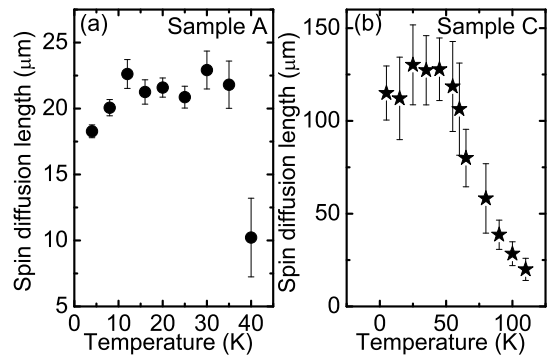}
   \caption{ (a) Spin diffusion length as a function of temperature for samples A(a) and C(b).}
   \label{Diff_2Panel}
\end{figure}
\begin{table}
\begin{tabular}
{|l|c|c|c|c|c|c|c|c|c|}
\hline
\# &  $\Omega_D$     &  $v$ & Temp. & $L_z$ & $\tau_s$ & $\tau_z$  & $\tau_{||}$ & $D_z$  \\
 & $10^{10}$s$^{-1}$       &  $10^{5}$ms$^{-1}$ & K & $\mu$m & ns & ns & ns & cm$^2$s$^{-1}$ \\
\hline
A           & 2.9     & 2.2   & 35 & 22 & 10 & 28 & 3.5  & 171  \\     
\hline
C          & 5.6      & 1.9    & 55 & 120 & 15.1   & 523 & 0.4 & 275  \\ 
\hline
\end{tabular}
\caption{ Experimental data and parameters obtained for sample A and C. The frequency $\Omega_D$ is calculated for the bulk Dresselhaus coefficient $\gamma \sim 12$\,eV$\cdot$\AA$^3$, which was determined experimentally for QWs with similar confinement energies as our samples~\cite{Eldridge11}. $D_z$, $\tau_z$, and $\tau_{\parallel}$ are calculated following Eqs.~(\ref{tau_z_est})-(\ref{D_z_est}).}
\label{Diffdata}
\end{table}

Finally, we study the temperature dependence of the spin diffusion length. Figure ~\ref{Diff_2Panel} shows the diffusion length in samples A and C as a function of temperature. We note that the spin diffusion length in sample C is more than five times  larger than that in sample A at low temperatures, and that spin diffusion can be studied in sample C for temperatures above 100~K, while the Hanle-MOKE signal in sample A is lost above 40~K. Additionally, both samples show a near-constant value of the spin diffusion length in a large temperature window. This is remarkable, given that the spin lifetime in sample A drops monotonously from 40~ns at 4~K to about 10~ns at 30~K (Fig.~\ref{Tspin_Temp}(a)), while the spin lifetime in sample C shows a pronounced maximum at 18~K and varies by a factor of 4 in the temperature window between 4~K and 50~K (Fig.~\ref{Tspin_Temp}(b)), in measurements with a large focal spot.  From the constant spin diffusion length we infer that the electron gas is significantly overheated at sample temperatures below 30~K due to the large excitation density resulting from the tightly focussed excitation, so that changes in the lattice temperature do not modify  spin dephasing time nor diffusion coefficient. Remarkably, this overheating apparently influences the spin diffusion also at large distances from the pump spot.   Overheating effects were recently studied in low-doped bulk GaAs, and shown to be relevant  at temperatures below 25~K and distances of up to 30~$\mu$m~\cite{Ossau13}.  They can be even more pronounced also at higher temperatures due to the high mobility in our 2DES: it corresponds to a very weak coupling between the electron system and the lattice via inelastic scattering processes, so that the colder lattice is an inefficient heat sink for the electron system.  Using the assumption that the electron temperature during the diffusion measurements corresponds to the highest sample temperature for which the diffusion length is constant, we can extract values for the spin diffusion coefficient $D_z$, out-of-plane and in-plane spin dephasing times, $\tau_z$ and $\tau_{\parallel}$, respectively, following Eqs.~(\ref{tau_z_est})-(\ref{D_z_est}). The obtained values are summarized in table~\ref{Diffdata}.
We find $D_z$(sample A, 35~K)=171~cm$^2$s$^{-1}$ and $D_z$(sample C, 55~K)=275~cm$^2$s$^{-1}$, far below  the values for the charge diffusion coefficient $D_n$ obtained from magnetotransport data at lower temperatures, but larger than the  $D_z^{({\rm ee})}$   values calculated for the  degenerate electron gas with the estimated experimental temperature using equation~(\ref{Diff_spin}).  While these low values of $D_z$ limit the spin diffusion length, they may in fact be beneficial in spin transport devices in which an electric field is used to laterally drag packets of spin-polarized carriers. In this case, the spin-Coulomb-drag can considerably suppress spatial dispersion of the spin packets.

\section*{Conclusion}
In conclusion, we have investigated spin dephasing, spin polarization and spin diffusion in a series of high-mobility (110)-grown 2DES. In the limit of weak excitation, we observe long spin lifetimes above 100~ns in our samples at low temperatures. With increasing excitation density, the spin lifetimes are reduced due to additional dephasing induced by the photogenerated hole density, namely the BAP mechanism, and  recombination of spin-polarized electrons with holes. The spin dynamics in the samples are strongly influenced by the carrier density and  profile of the ionization of remote donors, which can be controlled by temperature and above-barrier illumination. The absolute value of the average spin polarization  is estimated from photoluminescence data collected under circularly polarized excitation and  exceeds 5~percent.  Due to the degenerate distribution of electrons in the samples, photoluminescence-based studies  give only limited access to  the spin dynamics. Spin diffusion profiles show diffusion lengths of more than 100~$\mu$m and a nonmonotonic dependence of the average spin polarization on the pump-probe distance for high excitation densities. The temperature-dependent spin diffusion measurements indicate that the 2DES is significantly overheated under focused photoexcitation for lattice temperatures below 30~K.

\section*{Acknowledgements}
The authors would like to thank E.L. Ivchenko, M.M. Glazov, J.-H. Quast, T. Henn, T. Kiessling, W. Ossau, B. Endres and E.Ya. Sherman for fruitful discussion.  Financial support by the DFG via SPP 1285 and SFB 689, RFBR,  and RF president grant MD-2062.2012.2  is gratefully acknowledged.

\bibliography{RSA110}

\begin{thebibliography}{51}
\expandafter\ifx\csname natexlab\endcsname\relax\def\natexlab#1{#1}\fi
\expandafter\ifx\csname bibnamefont\endcsname\relax
  \def\bibnamefont#1{#1}\fi
\expandafter\ifx\csname bibfnamefont\endcsname\relax
  \def\bibfnamefont#1{#1}\fi
\expandafter\ifx\csname citenamefont\endcsname\relax
  \def\citenamefont#1{#1}\fi
\expandafter\ifx\csname url\endcsname\relax
  \def\url#1{\texttt{#1}}\fi
\expandafter\ifx\csname urlprefix\endcsname\relax\def\urlprefix{URL }\fi
\providecommand{\bibinfo}[2]{#2}
\providecommand{\eprint}[2][]{\url{#2}}

\bibitem[{\citenamefont{Awschalom et~al.}(2002)\citenamefont{Awschalom, Loss,
  and Samarth}}]{awschalom02}
\bibinfo{author}{\bibfnamefont{D.}~\bibnamefont{Awschalom}},
  \bibinfo{author}{\bibfnamefont{D.}~\bibnamefont{Loss}}, \bibnamefont{and}
  \bibinfo{author}{\bibfnamefont{N.}~\bibnamefont{Samarth}},
  \emph{\bibinfo{title}{{Semiconductor Spintronics and Quantum Computation}}}
  (\bibinfo{publisher}{Springer}, \bibinfo{year}{2002}).

\bibitem[{\citenamefont{Fabian et~al.}(2007)\citenamefont{Fabian,
  Matos-Abiague, Ertler, Stano, and Zutic}}]{fabian07}
\bibinfo{author}{\bibfnamefont{J.}~\bibnamefont{Fabian}},
  \bibinfo{author}{\bibfnamefont{A.}~\bibnamefont{Matos-Abiague}},
  \bibinfo{author}{\bibfnamefont{C.}~\bibnamefont{Ertler}},
  \bibinfo{author}{\bibfnamefont{P.}~\bibnamefont{Stano}}, \bibnamefont{and}
  \bibinfo{author}{\bibfnamefont{I.}~\bibnamefont{Zutic}},
  \bibinfo{journal}{Acta Physica Slovaca} \textbf{\bibinfo{volume}{57}},
  \bibinfo{pages}{565} (\bibinfo{year}{2007}).

\bibitem[{\citenamefont{Wu et~al.}(2010)\citenamefont{Wu, Jiang, and
  Weng}}]{WuReview}
\bibinfo{author}{\bibfnamefont{M.~W.} \bibnamefont{Wu}},
  \bibinfo{author}{\bibfnamefont{J.~H.} \bibnamefont{Jiang}}, \bibnamefont{and}
  \bibinfo{author}{\bibfnamefont{M.~Q.} \bibnamefont{Weng}},
  \bibinfo{journal}{Physics Reports} \textbf{\bibinfo{volume}{493}},
  \bibinfo{pages}{61} (\bibinfo{year}{2010}).

\bibitem[{\citenamefont{Datta and Das}(1990)}]{DattaDas}
\bibinfo{author}{\bibfnamefont{S.}~\bibnamefont{Datta}} \bibnamefont{and}
  \bibinfo{author}{\bibfnamefont{B.}~\bibnamefont{Das}},
  \bibinfo{journal}{Appl. Phys. Lett.} \textbf{\bibinfo{volume}{56}},
  \bibinfo{pages}{665} (\bibinfo{year}{1990}).

\bibitem[{\citenamefont{Schliemann et~al.}(2003)\citenamefont{Schliemann,
  Egues, and Loss}}]{Schliemann03}
\bibinfo{author}{\bibfnamefont{J.}~\bibnamefont{Schliemann}},
  \bibinfo{author}{\bibfnamefont{J.~C.} \bibnamefont{Egues}}, \bibnamefont{and}
  \bibinfo{author}{\bibfnamefont{D.}~\bibnamefont{Loss}},
  \bibinfo{journal}{Phys. Rev. Lett.} \textbf{\bibinfo{volume}{90}},
  \bibinfo{pages}{146801} (\bibinfo{year}{2003}).

\bibitem[{\citenamefont{Kunihashi et~al.}(2012)\citenamefont{Kunihashi, Kohda,
  Sanada, Gotoh, Sogawa, and Nitta}}]{kunihashi:113502}
\bibinfo{author}{\bibfnamefont{Y.}~\bibnamefont{Kunihashi}},
  \bibinfo{author}{\bibfnamefont{M.}~\bibnamefont{Kohda}},
  \bibinfo{author}{\bibfnamefont{H.}~\bibnamefont{Sanada}},
  \bibinfo{author}{\bibfnamefont{H.}~\bibnamefont{Gotoh}},
  \bibinfo{author}{\bibfnamefont{T.}~\bibnamefont{Sogawa}}, \bibnamefont{and}
  \bibinfo{author}{\bibfnamefont{J.}~\bibnamefont{Nitta}},
  \bibinfo{journal}{Appl. Phys. Lett.} \textbf{\bibinfo{volume}{100}},
  \bibinfo{eid}{113502} (\bibinfo{year}{2012}).

\bibitem[{\citenamefont{Koo et~al.}(2009)\citenamefont{Koo, Kwon, Eom, Chang,
  Han, and Johnson}}]{Koo09182009}
\bibinfo{author}{\bibfnamefont{H.~C.} \bibnamefont{Koo}},
  \bibinfo{author}{\bibfnamefont{J.~H.} \bibnamefont{Kwon}},
  \bibinfo{author}{\bibfnamefont{J.}~\bibnamefont{Eom}},
  \bibinfo{author}{\bibfnamefont{J.}~\bibnamefont{Chang}},
  \bibinfo{author}{\bibfnamefont{S.~H.} \bibnamefont{Han}}, \bibnamefont{and}
  \bibinfo{author}{\bibfnamefont{M.}~\bibnamefont{Johnson}},
  \bibinfo{journal}{Science} \textbf{\bibinfo{volume}{325}},
  \bibinfo{pages}{1515} (\bibinfo{year}{2009}).

\bibitem[{\citenamefont{Rudolph et~al.}(2003)\citenamefont{Rudolph, H\"{a}gele,
  Gibbs, Khitrova, and Oestreich}}]{Oestreich_SpinLaser}
\bibinfo{author}{\bibfnamefont{J.}~\bibnamefont{Rudolph}},
  \bibinfo{author}{\bibfnamefont{D.}~\bibnamefont{H\"{a}gele}},
  \bibinfo{author}{\bibfnamefont{H.~M.} \bibnamefont{Gibbs}},
  \bibinfo{author}{\bibfnamefont{G.}~\bibnamefont{Khitrova}}, \bibnamefont{and}
  \bibinfo{author}{\bibfnamefont{M.}~\bibnamefont{Oestreich}},
  \bibinfo{journal}{Appl. Phys. Lett.} \textbf{\bibinfo{volume}{82}},
  \bibinfo{pages}{4516} (\bibinfo{year}{2003}).

\bibitem[{\citenamefont{Lee et~al.}(2010)\citenamefont{Lee, Falls,
  Oszwa{\l}dowski, and \v{Z}uti\'{c}}}]{Zutic_SpinLaser}
\bibinfo{author}{\bibfnamefont{J.}~\bibnamefont{Lee}},
  \bibinfo{author}{\bibfnamefont{W.}~\bibnamefont{Falls}},
  \bibinfo{author}{\bibfnamefont{R.}~\bibnamefont{Oszwa{\l}dowski}},
  \bibnamefont{and}
  \bibinfo{author}{\bibfnamefont{I.}~\bibnamefont{\v{Z}uti\'{c}}},
  \bibinfo{journal}{Appl. Phys. Lett.} \textbf{\bibinfo{volume}{97}},
  \bibinfo{eid}{041116} (\bibinfo{year}{2010}).

\bibitem[{\citenamefont{Gerhardt et~al.}(2011)\citenamefont{Gerhardt, Li,
  J\"{a}hme, H\"{o}pfner, Ackemann, and Hofmann}}]{Hofmann_SpinLaser}
\bibinfo{author}{\bibfnamefont{N.~C.} \bibnamefont{Gerhardt}},
  \bibinfo{author}{\bibfnamefont{M.~Y.} \bibnamefont{Li}},
  \bibinfo{author}{\bibfnamefont{H.}~\bibnamefont{J\"{a}hme}},
  \bibinfo{author}{\bibfnamefont{H.}~\bibnamefont{H\"{o}pfner}},
  \bibinfo{author}{\bibfnamefont{T.}~\bibnamefont{Ackemann}}, \bibnamefont{and}
  \bibinfo{author}{\bibfnamefont{M.~R.} \bibnamefont{Hofmann}},
  \bibinfo{journal}{Appl. Phys. Lett.} \textbf{\bibinfo{volume}{99}},
  \bibinfo{eid}{151107} (\bibinfo{year}{2011}).

\bibitem[{\citenamefont{Giri et~al.}(2012)\citenamefont{Giri, Cronenberger,
  Vladimirova, Scalbert, Kavokin, Glazov, Nawrocki, Lemaitre, and
  Bloch}}]{Bloch_GiantFaraday}
\bibinfo{author}{\bibfnamefont{R.}~\bibnamefont{Giri}},
  \bibinfo{author}{\bibfnamefont{S.}~\bibnamefont{Cronenberger}},
  \bibinfo{author}{\bibfnamefont{M.}~\bibnamefont{Vladimirova}},
  \bibinfo{author}{\bibfnamefont{D.}~\bibnamefont{Scalbert}},
  \bibinfo{author}{\bibfnamefont{K.~V.} \bibnamefont{Kavokin}},
  \bibinfo{author}{\bibfnamefont{M.~M.} \bibnamefont{Glazov}},
  \bibinfo{author}{\bibfnamefont{M.}~\bibnamefont{Nawrocki}},
  \bibinfo{author}{\bibfnamefont{A.}~\bibnamefont{Lemaitre}}, \bibnamefont{and}
  \bibinfo{author}{\bibfnamefont{J.}~\bibnamefont{Bloch}},
  \bibinfo{journal}{Phys. Rev. B} \textbf{\bibinfo{volume}{85}},
  \bibinfo{pages}{195313} (\bibinfo{year}{2012}).

\bibitem[{\citenamefont{Dyakonov and Perel}({1971})}]{DP}
\bibinfo{author}{\bibfnamefont{M.}~\bibnamefont{Dyakonov}} \bibnamefont{and}
  \bibinfo{author}{\bibfnamefont{V.}~\bibnamefont{Perel}},
  \bibinfo{journal}{{Sov. Phys. JETP}} \textbf{\bibinfo{volume}{{33}}},
  \bibinfo{pages}{{1053}} (\bibinfo{year}{{1971}}).

\bibitem[{\citenamefont{Dyakonov and Kachorovskii}(1986)}]{DP110}
\bibinfo{author}{\bibfnamefont{M.}~\bibnamefont{Dyakonov}} \bibnamefont{and}
  \bibinfo{author}{\bibfnamefont{V.}~\bibnamefont{Kachorovskii}},
  \bibinfo{journal}{Sov. Phys. Semicond.} \textbf{\bibinfo{volume}{20}},
  \bibinfo{pages}{110} (\bibinfo{year}{1986}).

\bibitem[{\citenamefont{Tarasenko}(2009)}]{tarasenko:165317}
\bibinfo{author}{\bibfnamefont{S.~A.} \bibnamefont{Tarasenko}},
  \bibinfo{journal}{Phys. Rev. B} \textbf{\bibinfo{volume}{80}},
  \bibinfo{eid}{165317} (\bibinfo{year}{2009}).

\bibitem[{\citenamefont{Sherman}(2003)}]{Sherman_RandomRashba}
\bibinfo{author}{\bibfnamefont{E.~Y.} \bibnamefont{Sherman}},
  \bibinfo{journal}{Appl. Phys. Lett.} \textbf{\bibinfo{volume}{82}},
  \bibinfo{pages}{209} (\bibinfo{year}{2003}).

\bibitem[{\citenamefont{Poshakinskiy and Tarasenko}(2013)}]{Tarasenko13}
\bibinfo{author}{\bibfnamefont{A.~V.} \bibnamefont{Poshakinskiy}}
  \bibnamefont{and} \bibinfo{author}{\bibfnamefont{S.~A.}
  \bibnamefont{Tarasenko}}, \bibinfo{journal}{Phys. Rev. B}
  \textbf{\bibinfo{volume}{87}}, \bibinfo{pages}{235301}
  (\bibinfo{year}{2013}).

\bibitem[{\citenamefont{Ohno et~al.}(1999)\citenamefont{Ohno, Terauchi, Adachi,
  Matsukura, and Ohno}}]{Ohno99_1}
\bibinfo{author}{\bibfnamefont{Y.}~\bibnamefont{Ohno}},
  \bibinfo{author}{\bibfnamefont{R.}~\bibnamefont{Terauchi}},
  \bibinfo{author}{\bibfnamefont{T.}~\bibnamefont{Adachi}},
  \bibinfo{author}{\bibfnamefont{F.}~\bibnamefont{Matsukura}},
  \bibnamefont{and} \bibinfo{author}{\bibfnamefont{H.}~\bibnamefont{Ohno}},
  \bibinfo{journal}{Phys. Rev. Lett.} \textbf{\bibinfo{volume}{83}},
  \bibinfo{pages}{4196} (\bibinfo{year}{1999}).

\bibitem[{\citenamefont{Salis et~al.}(2001)\citenamefont{Salis, Fuchs, Kikkawa,
  Awschalom, Ohno, and Ohno}}]{ohno01}
\bibinfo{author}{\bibfnamefont{G.}~\bibnamefont{Salis}},
  \bibinfo{author}{\bibfnamefont{D.~T.} \bibnamefont{Fuchs}},
  \bibinfo{author}{\bibfnamefont{J.~M.} \bibnamefont{Kikkawa}},
  \bibinfo{author}{\bibfnamefont{D.~D.} \bibnamefont{Awschalom}},
  \bibinfo{author}{\bibfnamefont{Y.}~\bibnamefont{Ohno}}, \bibnamefont{and}
  \bibinfo{author}{\bibfnamefont{H.}~\bibnamefont{Ohno}},
  \bibinfo{journal}{Phys. Rev. Lett.} \textbf{\bibinfo{volume}{86}},
  \bibinfo{pages}{2677} (\bibinfo{year}{2001}).

\bibitem[{\citenamefont{D\"{o}hrmann et~al.}(2004)\citenamefont{D\"{o}hrmann,
  H\"{a}gele, Rudolph, Bichler, Schuh, and Oestreich}}]{Oestreich04_1}
\bibinfo{author}{\bibfnamefont{S.}~\bibnamefont{D\"{o}hrmann}},
  \bibinfo{author}{\bibfnamefont{D.}~\bibnamefont{H\"{a}gele}},
  \bibinfo{author}{\bibfnamefont{J.}~\bibnamefont{Rudolph}},
  \bibinfo{author}{\bibfnamefont{M.}~\bibnamefont{Bichler}},
  \bibinfo{author}{\bibfnamefont{D.}~\bibnamefont{Schuh}}, \bibnamefont{and}
  \bibinfo{author}{\bibfnamefont{M.}~\bibnamefont{Oestreich}},
  \bibinfo{journal}{Phys. Rev. Lett.} \textbf{\bibinfo{volume}{93}},
  \bibinfo{eid}{147405} (\bibinfo{year}{2004}).

\bibitem[{\citenamefont{M\"{u}ller et~al.}(2008)\citenamefont{M\"{u}ller,
  R\"{o}mer, Schuh, Wegscheider, H\"{u}bner, and Oestreich}}]{oestreich08}
\bibinfo{author}{\bibfnamefont{G.~M.} \bibnamefont{M\"{u}ller}},
  \bibinfo{author}{\bibfnamefont{M.}~\bibnamefont{R\"{o}mer}},
  \bibinfo{author}{\bibfnamefont{D.}~\bibnamefont{Schuh}},
  \bibinfo{author}{\bibfnamefont{W.}~\bibnamefont{Wegscheider}},
  \bibinfo{author}{\bibfnamefont{J.}~\bibnamefont{H\"{u}bner}},
  \bibnamefont{and}
  \bibinfo{author}{\bibfnamefont{M.}~\bibnamefont{Oestreich}},
  \bibinfo{journal}{Phys. Rev. Lett.} \textbf{\bibinfo{volume}{101}},
  \bibinfo{eid}{206601} (\bibinfo{year}{2008}).

\bibitem[{\citenamefont{V\"olkl et~al.}(2011)\citenamefont{V\"olkl, Griesbeck,
  Tarasenko, Schuh, Wegscheider, Sch\"uller, and Korn}}]{Voelkl11}
\bibinfo{author}{\bibfnamefont{R.}~\bibnamefont{V\"olkl}},
  \bibinfo{author}{\bibfnamefont{M.}~\bibnamefont{Griesbeck}},
  \bibinfo{author}{\bibfnamefont{S.~A.} \bibnamefont{Tarasenko}},
  \bibinfo{author}{\bibfnamefont{D.}~\bibnamefont{Schuh}},
  \bibinfo{author}{\bibfnamefont{W.}~\bibnamefont{Wegscheider}},
  \bibinfo{author}{\bibfnamefont{C.}~\bibnamefont{Sch\"uller}},
  \bibnamefont{and} \bibinfo{author}{\bibfnamefont{T.}~\bibnamefont{Korn}},
  \bibinfo{journal}{Phys. Rev. B} \textbf{\bibinfo{volume}{83}},
  \bibinfo{pages}{241306} (\bibinfo{year}{2011}).

\bibitem[{\citenamefont{Griesbeck et~al.}(2012)\citenamefont{Griesbeck, Glazov,
  Sherman, Schuh, Wegscheider, Sch\"uller, and Korn}}]{Griesbeck12}
\bibinfo{author}{\bibfnamefont{M.}~\bibnamefont{Griesbeck}},
  \bibinfo{author}{\bibfnamefont{M.~M.} \bibnamefont{Glazov}},
  \bibinfo{author}{\bibfnamefont{E.~Y.} \bibnamefont{Sherman}},
  \bibinfo{author}{\bibfnamefont{D.}~\bibnamefont{Schuh}},
  \bibinfo{author}{\bibfnamefont{W.}~\bibnamefont{Wegscheider}},
  \bibinfo{author}{\bibfnamefont{C.}~\bibnamefont{Sch\"uller}},
  \bibnamefont{and} \bibinfo{author}{\bibfnamefont{T.}~\bibnamefont{Korn}},
  \bibinfo{journal}{Phys. Rev. B} \textbf{\bibinfo{volume}{85}},
  \bibinfo{pages}{085313} (\bibinfo{year}{2012}).

\bibitem[{\citenamefont{Kikkawa and Awschalom}(1999)}]{awschalom99}
\bibinfo{author}{\bibfnamefont{J.}~\bibnamefont{Kikkawa}} \bibnamefont{and}
  \bibinfo{author}{\bibfnamefont{D.}~\bibnamefont{Awschalom}},
  \bibinfo{journal}{Nature} \textbf{\bibinfo{volume}{397}},
  \bibinfo{pages}{139} (\bibinfo{year}{1999}).

\bibitem[{\citenamefont{Crooker and Smith}(2005)}]{crooker05_1}
\bibinfo{author}{\bibfnamefont{S.~A.} \bibnamefont{Crooker}} \bibnamefont{and}
  \bibinfo{author}{\bibfnamefont{D.~L.} \bibnamefont{Smith}},
  \bibinfo{journal}{Phys. Rev. Lett.} \textbf{\bibinfo{volume}{94}},
  \bibinfo{eid}{236601} (\bibinfo{year}{2005}).

\bibitem[{\citenamefont{Crooker et~al.}(2005)\citenamefont{Crooker, Furis, Lou,
  Adelmann, Smith, Palmstrom, and Crowell}}]{crooker05_2}
\bibinfo{author}{\bibfnamefont{S.~A.} \bibnamefont{Crooker}},
  \bibinfo{author}{\bibfnamefont{M.}~\bibnamefont{Furis}},
  \bibinfo{author}{\bibfnamefont{X.}~\bibnamefont{Lou}},
  \bibinfo{author}{\bibfnamefont{C.}~\bibnamefont{Adelmann}},
  \bibinfo{author}{\bibfnamefont{D.~L.} \bibnamefont{Smith}},
  \bibinfo{author}{\bibfnamefont{C.~J.} \bibnamefont{Palmstrom}},
  \bibnamefont{and} \bibinfo{author}{\bibfnamefont{P.~A.}
  \bibnamefont{Crowell}}, \bibinfo{journal}{Science}
  \textbf{\bibinfo{volume}{309}}, \bibinfo{pages}{2191} (\bibinfo{year}{2005}).

\bibitem[{\citenamefont{Furis et~al.}(2007)\citenamefont{Furis, Smith, Kos,
  Garlid, Reddy, Palmstr\o{}m, Crowell, and Crooker}}]{crooker07}
\bibinfo{author}{\bibfnamefont{M.}~\bibnamefont{Furis}},
  \bibinfo{author}{\bibfnamefont{D.~L.} \bibnamefont{Smith}},
  \bibinfo{author}{\bibfnamefont{S.}~\bibnamefont{Kos}},
  \bibinfo{author}{\bibfnamefont{E.~S.} \bibnamefont{Garlid}},
  \bibinfo{author}{\bibfnamefont{K.~S.~M.} \bibnamefont{Reddy}},
  \bibinfo{author}{\bibfnamefont{C.~J.} \bibnamefont{Palmstr\o{}m}},
  \bibinfo{author}{\bibfnamefont{P.~A.} \bibnamefont{Crowell}},
  \bibnamefont{and} \bibinfo{author}{\bibfnamefont{S.~A.}
  \bibnamefont{Crooker}}, \bibinfo{journal}{New J. Phys.}
  \textbf{\bibinfo{volume}{9}}, \bibinfo{pages}{347} (\bibinfo{year}{2007}).

\bibitem[{\citenamefont{Quast et~al.}(2009)\citenamefont{Quast, Astakhov,
  Ossau, Molenkamp, Heinrich, H\"ofling, and Forchel}}]{Ossau09}
\bibinfo{author}{\bibfnamefont{J.-H.} \bibnamefont{Quast}},
  \bibinfo{author}{\bibfnamefont{G.~V.} \bibnamefont{Astakhov}},
  \bibinfo{author}{\bibfnamefont{W.}~\bibnamefont{Ossau}},
  \bibinfo{author}{\bibfnamefont{L.~W.} \bibnamefont{Molenkamp}},
  \bibinfo{author}{\bibfnamefont{J.}~\bibnamefont{Heinrich}},
  \bibinfo{author}{\bibfnamefont{S.}~\bibnamefont{H\"ofling}},
  \bibnamefont{and} \bibinfo{author}{\bibfnamefont{A.}~\bibnamefont{Forchel}},
  \bibinfo{journal}{Phys. Rev. B} \textbf{\bibinfo{volume}{79}},
  \bibinfo{pages}{245207} (\bibinfo{year}{2009}).

\bibitem[{\citenamefont{Cameron et~al.}(1996)\citenamefont{Cameron, Riblet, and
  Miller}}]{SpinGrating96}
\bibinfo{author}{\bibfnamefont{A.~R.} \bibnamefont{Cameron}},
  \bibinfo{author}{\bibfnamefont{P.}~\bibnamefont{Riblet}}, \bibnamefont{and}
  \bibinfo{author}{\bibfnamefont{A.}~\bibnamefont{Miller}},
  \bibinfo{journal}{Phys. Rev. Lett.} \textbf{\bibinfo{volume}{76}},
  \bibinfo{pages}{4793} (\bibinfo{year}{1996}).

\bibitem[{\citenamefont{Chen et~al.}(2012)\citenamefont{Chen, Wang, Wu, Schuh,
  Wegscheider, Korn, and Lai}}]{Chen:12}
\bibinfo{author}{\bibfnamefont{K.}~\bibnamefont{Chen}},
  \bibinfo{author}{\bibfnamefont{W.}~\bibnamefont{Wang}},
  \bibinfo{author}{\bibfnamefont{J.}~\bibnamefont{Wu}},
  \bibinfo{author}{\bibfnamefont{D.}~\bibnamefont{Schuh}},
  \bibinfo{author}{\bibfnamefont{W.}~\bibnamefont{Wegscheider}},
  \bibinfo{author}{\bibfnamefont{T.}~\bibnamefont{Korn}}, \bibnamefont{and}
  \bibinfo{author}{\bibfnamefont{T.}~\bibnamefont{Lai}}, \bibinfo{journal}{Opt.
  Express} \textbf{\bibinfo{volume}{20}}, \bibinfo{pages}{8192}
  (\bibinfo{year}{2012}).

\bibitem[{\citenamefont{Couto et~al.}(2007)\citenamefont{Couto, Iikawa,
  Rudolph, Hey, and Santos}}]{Santos_transport110_PRL07}
\bibinfo{author}{\bibfnamefont{O.~D.~D.} \bibnamefont{Couto}},
  \bibinfo{author}{\bibfnamefont{F.}~\bibnamefont{Iikawa}},
  \bibinfo{author}{\bibfnamefont{J.}~\bibnamefont{Rudolph}},
  \bibinfo{author}{\bibfnamefont{R.}~\bibnamefont{Hey}}, \bibnamefont{and}
  \bibinfo{author}{\bibfnamefont{P.~V.} \bibnamefont{Santos}},
  \bibinfo{journal}{Phys. Rev. Lett.} \textbf{\bibinfo{volume}{98}},
  \bibinfo{pages}{036603} (\bibinfo{year}{2007}).

\bibitem[{\citenamefont{Couto et~al.}(2008)\citenamefont{Couto, Hey, and
  Santos}}]{Santos_transport110_PRB08}
\bibinfo{author}{\bibfnamefont{O.~D.~D.} \bibnamefont{Couto}},
  \bibinfo{author}{\bibfnamefont{R.}~\bibnamefont{Hey}}, \bibnamefont{and}
  \bibinfo{author}{\bibfnamefont{P.~V.} \bibnamefont{Santos}},
  \bibinfo{journal}{Phys. Rev. B} \textbf{\bibinfo{volume}{78}},
  \bibinfo{pages}{153305} (\bibinfo{year}{2008}).

\bibitem[{\citenamefont{Hern\'{a}ndez-M\'{\i}nguez
  et~al.}(2010)\citenamefont{Hern\'{a}ndez-M\'{\i}nguez, Biermann, Lazi\'{c},
  Hey, and Santos}}]{Santos10}
\bibinfo{author}{\bibfnamefont{A.}~\bibnamefont{Hern\'{a}ndez-M\'{\i}nguez}},
  \bibinfo{author}{\bibfnamefont{K.}~\bibnamefont{Biermann}},
  \bibinfo{author}{\bibfnamefont{S.}~\bibnamefont{Lazi\'{c}}},
  \bibinfo{author}{\bibfnamefont{R.}~\bibnamefont{Hey}}, \bibnamefont{and}
  \bibinfo{author}{\bibfnamefont{P.~V.} \bibnamefont{Santos}},
  \bibinfo{journal}{Appl. Phys. Lett.} \textbf{\bibinfo{volume}{97}},
  \bibinfo{eid}{242110} (\bibinfo{year}{2010}).

\bibitem[{\citenamefont{Hu et~al.}(2011)\citenamefont{Hu, Ye, Wang, Tian, Wang,
  Wang, Liu, and Marie}}]{Marie11}
\bibinfo{author}{\bibfnamefont{C.}~\bibnamefont{Hu}},
  \bibinfo{author}{\bibfnamefont{H.}~\bibnamefont{Ye}},
  \bibinfo{author}{\bibfnamefont{G.}~\bibnamefont{Wang}},
  \bibinfo{author}{\bibfnamefont{H.}~\bibnamefont{Tian}},
  \bibinfo{author}{\bibfnamefont{W.}~\bibnamefont{Wang}},
  \bibinfo{author}{\bibfnamefont{W.}~\bibnamefont{Wang}},
  \bibinfo{author}{\bibfnamefont{B.}~\bibnamefont{Liu}}, \bibnamefont{and}
  \bibinfo{author}{\bibfnamefont{X.}~\bibnamefont{Marie}},
  \bibinfo{journal}{Nanoscale Res. Lett.} \textbf{\bibinfo{volume}{6}},
  \bibinfo{pages}{149} (\bibinfo{year}{2011}).

\bibitem[{\citenamefont{Umansky et~al.}(2009)\citenamefont{Umansky, Heiblum,
  Levinson, Smet, N\"ubler, and Dolev}}]{Umansky20091658}
\bibinfo{author}{\bibfnamefont{V.}~\bibnamefont{Umansky}},
  \bibinfo{author}{\bibfnamefont{M.}~\bibnamefont{Heiblum}},
  \bibinfo{author}{\bibfnamefont{Y.}~\bibnamefont{Levinson}},
  \bibinfo{author}{\bibfnamefont{J.}~\bibnamefont{Smet}},
  \bibinfo{author}{\bibfnamefont{J.}~\bibnamefont{N\"ubler}}, \bibnamefont{and}
  \bibinfo{author}{\bibfnamefont{M.}~\bibnamefont{Dolev}}, \bibinfo{journal}{J.
  Crystal Growth} \textbf{\bibinfo{volume}{311}}, \bibinfo{pages}{1658}
  (\bibinfo{year}{2009}).

\bibitem[{\citenamefont{Pfalz et~al.}(2005)\citenamefont{Pfalz, Winkler,
  Nowitzki, Reuter, Wieck, H\"{a}gele, and Oestreich}}]{oestreich05_1}
\bibinfo{author}{\bibfnamefont{S.}~\bibnamefont{Pfalz}},
  \bibinfo{author}{\bibfnamefont{R.}~\bibnamefont{Winkler}},
  \bibinfo{author}{\bibfnamefont{T.}~\bibnamefont{Nowitzki}},
  \bibinfo{author}{\bibfnamefont{D.}~\bibnamefont{Reuter}},
  \bibinfo{author}{\bibfnamefont{A.~D.} \bibnamefont{Wieck}},
  \bibinfo{author}{\bibfnamefont{D.}~\bibnamefont{H\"{a}gele}},
  \bibnamefont{and}
  \bibinfo{author}{\bibfnamefont{M.}~\bibnamefont{Oestreich}},
  \bibinfo{journal}{Phys. Rev. B} \textbf{\bibinfo{volume}{71}},
  \bibinfo{eid}{165305} (\bibinfo{year}{2005}).

\bibitem[{\citenamefont{Zhou and Wu}(2010)}]{Wu_randomRashba}
\bibinfo{author}{\bibfnamefont{Y.}~\bibnamefont{Zhou}} \bibnamefont{and}
  \bibinfo{author}{\bibfnamefont{M.~W.} \bibnamefont{Wu}},
  \bibinfo{journal}{EPL (Europhysics Letters)} \textbf{\bibinfo{volume}{89}},
  \bibinfo{pages}{57001} (\bibinfo{year}{2010}).

\bibitem[{\citenamefont{Bir et~al.}(1975)\citenamefont{Bir, Aronov, and
  Pikus}}]{BAP}
\bibinfo{author}{\bibfnamefont{G.~L.} \bibnamefont{Bir}},
  \bibinfo{author}{\bibfnamefont{A.~G.} \bibnamefont{Aronov}},
  \bibnamefont{and} \bibinfo{author}{\bibfnamefont{G.~E.} \bibnamefont{Pikus}},
  \bibinfo{journal}{Sov. Phys. JETP} \textbf{\bibinfo{volume}{42}},
  \bibinfo{pages}{705} (\bibinfo{year}{1975}).

\bibitem[{\citenamefont{Aronov et~al.}(1983)\citenamefont{Aronov, Pikus, and
  Titkov}}]{APT83}
\bibinfo{author}{\bibfnamefont{A.~G.} \bibnamefont{Aronov}},
  \bibinfo{author}{\bibfnamefont{G.~E.} \bibnamefont{Pikus}}, \bibnamefont{and}
  \bibinfo{author}{\bibfnamefont{A.~N.} \bibnamefont{Titkov}},
  \bibinfo{journal}{Zh. Eksp. Teor. Fiz.} \textbf{\bibinfo{volume}{84}},
  \bibinfo{pages}{1170} (\bibinfo{year}{1983}).

\bibitem[{\citenamefont{Leyland et~al.}(2007)\citenamefont{Leyland, John,
  Harley, Glazov, Ivchenko, Ritchie, Farrer, Shields, and Henini}}]{Harley07}
\bibinfo{author}{\bibfnamefont{W.~J.~H.} \bibnamefont{Leyland}},
  \bibinfo{author}{\bibfnamefont{G.~H.} \bibnamefont{John}},
  \bibinfo{author}{\bibfnamefont{R.~T.} \bibnamefont{Harley}},
  \bibinfo{author}{\bibfnamefont{M.~M.} \bibnamefont{Glazov}},
  \bibinfo{author}{\bibfnamefont{E.~L.} \bibnamefont{Ivchenko}},
  \bibinfo{author}{\bibfnamefont{D.~A.} \bibnamefont{Ritchie}},
  \bibinfo{author}{\bibfnamefont{I.}~\bibnamefont{Farrer}},
  \bibinfo{author}{\bibfnamefont{A.~J.} \bibnamefont{Shields}},
  \bibnamefont{and} \bibinfo{author}{\bibfnamefont{M.}~\bibnamefont{Henini}},
  \bibinfo{journal}{Phys. Rev. B} \textbf{\bibinfo{volume}{75}},
  \bibinfo{eid}{165309} (\bibinfo{year}{2007}).

\bibitem[{\citenamefont{D'Amico and Vignale}(2000)}]{Damico00}
\bibinfo{author}{\bibfnamefont{I.}~\bibnamefont{D'Amico}} \bibnamefont{and}
  \bibinfo{author}{\bibfnamefont{G.}~\bibnamefont{Vignale}},
  \bibinfo{journal}{Phys. Rev. B} \textbf{\bibinfo{volume}{62}},
  \bibinfo{pages}{4853} (\bibinfo{year}{2000}).

\bibitem[{\citenamefont{Ivchenko and Tarasenko}(2008)}]{Ivchenko08}
\bibinfo{author}{\bibfnamefont{E.}~\bibnamefont{Ivchenko}} \bibnamefont{and}
  \bibinfo{author}{\bibfnamefont{S.}~\bibnamefont{Tarasenko}},
  \bibinfo{journal}{Semicond. Sci. Technol} \textbf{\bibinfo{volume}{23}},
  \bibinfo{pages}{114007} (\bibinfo{year}{2008}).

\bibitem[{\citenamefont{Brand et~al.}(2002)\citenamefont{Brand, Malinowski,
  Karimov, Marsden, Harley, Shields, Sanvitto, Ritchie, and
  Simmons}}]{Shields02}
\bibinfo{author}{\bibfnamefont{M.~A.} \bibnamefont{Brand}},
  \bibinfo{author}{\bibfnamefont{A.}~\bibnamefont{Malinowski}},
  \bibinfo{author}{\bibfnamefont{O.~Z.} \bibnamefont{Karimov}},
  \bibinfo{author}{\bibfnamefont{P.~A.} \bibnamefont{Marsden}},
  \bibinfo{author}{\bibfnamefont{R.~T.} \bibnamefont{Harley}},
  \bibinfo{author}{\bibfnamefont{A.~J.} \bibnamefont{Shields}},
  \bibinfo{author}{\bibfnamefont{D.}~\bibnamefont{Sanvitto}},
  \bibinfo{author}{\bibfnamefont{D.~A.} \bibnamefont{Ritchie}},
  \bibnamefont{and} \bibinfo{author}{\bibfnamefont{M.~Y.}
  \bibnamefont{Simmons}}, \bibinfo{journal}{Phys. Rev. Lett.}
  \textbf{\bibinfo{volume}{89}}, \bibinfo{pages}{236601}
  (\bibinfo{year}{2002}).

\bibitem[{\citenamefont{Griesbeck et~al.}(2009)\citenamefont{Griesbeck, Glazov,
  Korn, Sherman, Waller, Reichl, Schuh, Wegscheider, and
  Sch\"uller}}]{Griesbeck09}
\bibinfo{author}{\bibfnamefont{M.}~\bibnamefont{Griesbeck}},
  \bibinfo{author}{\bibfnamefont{M.~M.} \bibnamefont{Glazov}},
  \bibinfo{author}{\bibfnamefont{T.}~\bibnamefont{Korn}},
  \bibinfo{author}{\bibfnamefont{E.~Y.} \bibnamefont{Sherman}},
  \bibinfo{author}{\bibfnamefont{D.}~\bibnamefont{Waller}},
  \bibinfo{author}{\bibfnamefont{C.}~\bibnamefont{Reichl}},
  \bibinfo{author}{\bibfnamefont{D.}~\bibnamefont{Schuh}},
  \bibinfo{author}{\bibfnamefont{W.}~\bibnamefont{Wegscheider}},
  \bibnamefont{and}
  \bibinfo{author}{\bibfnamefont{C.}~\bibnamefont{Sch\"uller}},
  \bibinfo{journal}{Phys. Rev. B} \textbf{\bibinfo{volume}{80}},
  \bibinfo{pages}{241314} (\bibinfo{year}{2009}).

\bibitem[{\citenamefont{Zhou and Wu}(2008)}]{Zhou_Wu_BAP}
\bibinfo{author}{\bibfnamefont{J.}~\bibnamefont{Zhou}} \bibnamefont{and}
  \bibinfo{author}{\bibfnamefont{M.~W.} \bibnamefont{Wu}},
  \bibinfo{journal}{Phys. Rev. B} \textbf{\bibinfo{volume}{77}},
  \bibinfo{eid}{075318} (\bibinfo{year}{2008}).

\bibitem[{\citenamefont{Chaves et~al.}(1986)\citenamefont{Chaves, Penna,
  Worlock, Weimann, and Schlapp}}]{Chaves1986}
\bibinfo{author}{\bibfnamefont{A.}~\bibnamefont{Chaves}},
  \bibinfo{author}{\bibfnamefont{A.}~\bibnamefont{Penna}},
  \bibinfo{author}{\bibfnamefont{J.}~\bibnamefont{Worlock}},
  \bibinfo{author}{\bibfnamefont{G.}~\bibnamefont{Weimann}}, \bibnamefont{and}
  \bibinfo{author}{\bibfnamefont{W.}~\bibnamefont{Schlapp}},
  \bibinfo{journal}{Surface Science} \textbf{\bibinfo{volume}{170}},
  \bibinfo{pages}{618} (\bibinfo{year}{1986}).

\bibitem[{\citenamefont{Syperek et~al.}(2007)\citenamefont{Syperek, Yakovlev,
  Greilich, Misiewicz, Bayer, Reuter, and Wieck}}]{syperek07}
\bibinfo{author}{\bibfnamefont{M.}~\bibnamefont{Syperek}},
  \bibinfo{author}{\bibfnamefont{D.~R.} \bibnamefont{Yakovlev}},
  \bibinfo{author}{\bibfnamefont{A.}~\bibnamefont{Greilich}},
  \bibinfo{author}{\bibfnamefont{J.}~\bibnamefont{Misiewicz}},
  \bibinfo{author}{\bibfnamefont{M.}~\bibnamefont{Bayer}},
  \bibinfo{author}{\bibfnamefont{D.}~\bibnamefont{Reuter}}, \bibnamefont{and}
  \bibinfo{author}{\bibfnamefont{A.~D.} \bibnamefont{Wieck}},
  \bibinfo{journal}{Phys. Rev. Lett.} \textbf{\bibinfo{volume}{99}},
  \bibinfo{eid}{187401} (\bibinfo{year}{2007}).

\bibitem[{\citenamefont{Dzhioev et~al.}(2002)\citenamefont{Dzhioev, Kavokin,
  Korenev, Lazarev, Meltser, Stepanova, Zakharchenya, Gammon, and
  Katzer}}]{dzhioev02_2}
\bibinfo{author}{\bibfnamefont{R.}~\bibnamefont{Dzhioev}},
  \bibinfo{author}{\bibfnamefont{K.}~\bibnamefont{Kavokin}},
  \bibinfo{author}{\bibfnamefont{V.}~\bibnamefont{Korenev}},
  \bibinfo{author}{\bibfnamefont{M.}~\bibnamefont{Lazarev}},
  \bibinfo{author}{\bibfnamefont{B.}~\bibnamefont{Meltser}},
  \bibinfo{author}{\bibfnamefont{M.}~\bibnamefont{Stepanova}},
  \bibinfo{author}{\bibfnamefont{B.}~\bibnamefont{Zakharchenya}},
  \bibinfo{author}{\bibfnamefont{D.}~\bibnamefont{Gammon}}, \bibnamefont{and}
  \bibinfo{author}{\bibfnamefont{D.}~\bibnamefont{Katzer}},
  \bibinfo{journal}{Physical Review B} \textbf{\bibinfo{volume}{66}},
  \bibinfo{pages}{245204} (\bibinfo{year}{2002}).

\bibitem[{\citenamefont{Korn}(2010)}]{KornReview}
\bibinfo{author}{\bibfnamefont{T.}~\bibnamefont{Korn}},
  \bibinfo{journal}{Physics Reports} \textbf{\bibinfo{volume}{494}},
  \bibinfo{pages}{415} (\bibinfo{year}{2010}).

\bibitem[{\citenamefont{Kiessling et~al.}(2012)\citenamefont{Kiessling, Quast,
  Kreisel, Henn, Ossau, and Molenkamp}}]{Ossau12}
\bibinfo{author}{\bibfnamefont{T.}~\bibnamefont{Kiessling}},
  \bibinfo{author}{\bibfnamefont{J.-H.} \bibnamefont{Quast}},
  \bibinfo{author}{\bibfnamefont{A.}~\bibnamefont{Kreisel}},
  \bibinfo{author}{\bibfnamefont{T.}~\bibnamefont{Henn}},
  \bibinfo{author}{\bibfnamefont{W.}~\bibnamefont{Ossau}}, \bibnamefont{and}
  \bibinfo{author}{\bibfnamefont{L.~W.} \bibnamefont{Molenkamp}},
  \bibinfo{journal}{Phys. Rev. B} \textbf{\bibinfo{volume}{86}},
  \bibinfo{pages}{161201} (\bibinfo{year}{2012}).

\bibitem[{\citenamefont{Eldridge et~al.}(2011)\citenamefont{Eldridge, H\"ubner,
  Oertel, Harley, Henini, and Oestreich}}]{Eldridge11}
\bibinfo{author}{\bibfnamefont{P.~S.} \bibnamefont{Eldridge}},
  \bibinfo{author}{\bibfnamefont{J.}~\bibnamefont{H\"ubner}},
  \bibinfo{author}{\bibfnamefont{S.}~\bibnamefont{Oertel}},
  \bibinfo{author}{\bibfnamefont{R.~T.} \bibnamefont{Harley}},
  \bibinfo{author}{\bibfnamefont{M.}~\bibnamefont{Henini}}, \bibnamefont{and}
  \bibinfo{author}{\bibfnamefont{M.}~\bibnamefont{Oestreich}},
  \bibinfo{journal}{Phys. Rev. B} \textbf{\bibinfo{volume}{83}},
  \bibinfo{pages}{041301} (\bibinfo{year}{2011}).

\bibitem[{\citenamefont{Quast et~al.}(2013)\citenamefont{Quast, Henn,
  Kiessling, Ossau, Molenkamp, Reuter, and Wieck}}]{Ossau13}
\bibinfo{author}{\bibfnamefont{J.-H.} \bibnamefont{Quast}},
  \bibinfo{author}{\bibfnamefont{T.}~\bibnamefont{Henn}},
  \bibinfo{author}{\bibfnamefont{T.}~\bibnamefont{Kiessling}},
  \bibinfo{author}{\bibfnamefont{W.}~\bibnamefont{Ossau}},
  \bibinfo{author}{\bibfnamefont{L.~W.} \bibnamefont{Molenkamp}},
  \bibinfo{author}{\bibfnamefont{D.}~\bibnamefont{Reuter}}, \bibnamefont{and}
  \bibinfo{author}{\bibfnamefont{A.~D.} \bibnamefont{Wieck}},
  \bibinfo{journal}{Phys. Rev. B} \textbf{\bibinfo{volume}{87}},
  \bibinfo{pages}{205203} (\bibinfo{year}{2013}).

\end{thebibliography}

\end{document}